\documentclass[aps,amsfonts,draft]{revtex4}

\begin{document}

\title{How does the entropy$/$information bound work
?\footnote{Warmly dedicated to Asher Peres on the occasion of his
seventieth birthday.}}
\author{\bf Jacob D. Bekenstein\footnote{Racah Institute of Physics, The Hebrew University of Jerusalem, Givat Ram, Jerusalem 91904, Israel}}
\address{{}}
\date{\today}
\def\cu#1{\nabla\times{{\bf #1}}}
\def\cucu#1{\nabla\times\nabla\times{\bf #1}}
\def\di#1{\nabla\cdot{\bf #1}}
\def\ve#1{{\bf #1}}

\begin{abstract} According to the universal entropy bound, the
entropy (and hence information capacity) of a complete weakly
self-gravitating physical system can be  bounded exclusively in terms
of its circumscribing radius and total gravitating energy.   The bound's
correctness is supported by explicit statistical calculations of entropy,
gedanken experiments involving the generalized second law, and
Bousso's covariant holographic bound.   On the other hand, it is not
always obvious  in a particular example how the system avoids
having too many states for given energy, and hence violating the
bound.  We analyze in detail several purported counterexamples of
this type (involving systems made of massive particles, systems
at low temperature, systems with high degeneracy of the lowest
excited states, systems with degenerate ground states, or involving a
particle spectrum with proliferation of nearly massless species), and
exhibit in each case the mechanism behind the bound's efficacy. 

\end{abstract}

\maketitle

\section{Introduction}
\label{intro}

Information theory started as a theory of communication---transport
of information.  The developers of communication channel capacity
theorems paid little  attention to the akin question of information
\emph{storage} capacity. In essence such question boils down to a
more physically sounding one: what are the limitations on the
magnitude of the entropy of a system characterized by general
parameters such as size, energy, mass,
\dots ?   In 1981 I proposed~\cite{bek81} that the entropy of a
complete physical system in asymptotically flat
$D=4$ spacetime, whose total \emph{mass-energy} is
$E$, and which fits inside a sphere of radius $R$, is necessarily
bounded from above:
\begin{equation} S\leq 2\pi ER/\hbar c.
\label{bound}
\end{equation} The motivation for this
\emph{universal entropy bound} came from 
\emph{gedanken} experiments in which an entropy-bearing
object is deposited at a black hole's horizon with the least possible
energy; a violation of the generalized second law seems to occur
unless the said bound applies to the object~\cite{bek81}.   The
tenor of the argument is that
$E$ is to be interpreted as the gravitating energy of the system; this
prescription disposes of any ambiguity that would arise if we
attempted to redefine the zero of energy.

Unruh and Wald~\cite{UW} objected to the mentioned derivation by
pointing out the existence of quantum buoyancy of objects in a
black hole's vicinity.  Nevertheless, since quantum buoyancy is
significant only at distances from the horizon of the order of the
lowered object's size, the mentioned derivation can be suitably
amended to yield bound~(\ref{bound}) even in the face of
quantum buoyancy, except that a larger numerical coefficient
must be accepted~\cite{bek82,bek94}.  A variant of the
original \emph{gedanken} experiment~\cite{bek96,bek01} in which
the object is freely dropped into a black hole (and is thus immune to
quantum buoyancy) again gives bound~(\ref{bound}), albeit
with a larger coefficient.   Finally, Unruh and Wald's fluid model
of the quantum buoyancy has been shown to be
a gross approximation~\cite{bek99};  when the the responsible
radiation is treated as waves,  bound (\ref{bound}) can be
recovered even with buoyancy accounted for.  

Meanwhile 't Hooft~\cite{thooft} and Susskind~\cite{susskind}
introduced the holographic entropy bound
\begin{equation} S\leq \pi c^3 R^2/\hbar G,
\end{equation} 
where $G$ is the Newton's constant.  Again derivations from the generalized second
law~\cite{susskind,bek00} provided the clearest route.  It is now
understood that whereas the holographic bound is applicable to
all isolated physical systems, the universal bound is relevant only
for weakly self-gravitating isolated physical systems, and for these
it is a much stronger bound than the holographic
one~\cite{bek01}.   Bousso has shown how to derive the holographic
bound from his covariant entropy
bound~\cite{bousso99,boussorev}, which today is the most
generally applicable entropy bound known.  

A strengthened version of the \emph{universal} bound can be
obtained from the generalized covariant entropy
bound~\cite{bousso03,BFM} conjectured by Flanagan, Marolf and
Wald~\cite{FMW}.     But one
should  recall  that the  bound has its limitations.  These
principally belong to the strongly gravitating system regime.  In
common with the holographic bound, bound (\ref{bound}) does
not apply in wildly dynamic situations such as those found inside
black holes~\cite{bousso99}, and it is not guaranteed to work for
large pieces of the universe (which, after all, are not complete
systems).   Bound (\ref{bound}) does apply in higher
dimensions~\cite{bousso01,boussorev} and to entire closed
Robertson-Walker universes~\cite{verlinde}.  

There is no controversy today as to the validity of
bound~(\ref{bound}) for \emph{complete, weakly self-gravitating,
isolated objects in ordinary asymptotically flat spacetime}.  However,
the question of exactly how the universal entropy bound
(\ref{bound}) manages to sidestep various proposed counterexamples
to it has continued to be of interest for
two decades~\cite{page82,unwin,deutsch,bek82,bek83,unruh,BS,
bek94,page1,page2,page3}.    The present paper, which grew
from unpublished material~\cite{bekunp}, is devoted to an analysis of
several such attempts.  In each case it lays bare assumptions made by
the authors which contravene the conditions just mentioned for the
validity of the bound.  A paper by Bousso~\cite{bousso04} also deals
with the same issue, in some cases in a more detailed and precise
way.  The two papers are complementary.
 
\section{The rest mass quandary}
\label{sec:mass}

A counterexample frequently adduced against the universal entropy
bound by attentive listeners at lectures  can be elaborated as
follows.  Take a number $N\gg 1$ of nonrelativistic bosons of rest
mass $\mu$  confined to a space of typical  extent $R$.  If the
bosons can occupy $\Omega$ modes (one-particle states) in all,
the  number of states open to them is 
\begin{equation} W=(N+\Omega-1)!/[ (\Omega-1)! N!]
\end{equation} and the microcanonical entropy is $S=\ln W$. 
Now quantum mechanics tells us that the lowest lying modes have
energies
$\epsilon_0=\mathcal{O}(\hbar^2/\mu R^2)$ and this is also
the typical spacing between modes.    Therefore, our $N$ bosons,
if they are not in very excited states, will have a total energy
$E=\mathcal{O}(N\hbar^2/\mu R^2)$.   By making $\mu$
sufficiently large we make the entropy bound $2\pi ER/\hbar c$
so small that it will not be able to bound $S$.  Note that the
argument is one of scaling.   $S$ is unaffected by a rescaling of 
$\mu$.

What this argument glosses over is the stipulation that bound
(\ref{bound}) applies to a \emph{complete} system.  It leaves out
the contribution of rest energies $N\mu c^2$ to $E$, which after
all do gravitate.  By keeping the assumption of a nonrelativistic
system, we must interpret bound (\ref{bound}) as
\begin{equation} S<2\pi N\mu c R/\hbar
\label{bound1}
\end{equation}

To see if this is respected we approximate $\ln W$ with Sterling's
rule assuming not only $N\gg1$ but also $\Omega\gg1$.  Thus
\begin{eqnarray} S&=&(\Omega+N)\ln (\Omega+N)-N\ln
N-\Omega\ln\Omega+\cdots
\\ &=&
\Omega\ln(1+N/\Omega)+N\ln(1+\Omega/N)+\cdots
\label{ent}
\end{eqnarray} where  the ellipsis stand for corrections of order
$\ln N$ and
$\ln \Omega$ which are irrelevant to what follows.  Now the
number of available modes
$\Omega$, particularly when it is large as assumed,  is bounded
by the volume of phase space accessible to a nonrelativistic
particle.  Since a nonrelativistic particle's momentum must be
restricted to some small fraction of
$\mu c$ at most, we may write
\begin{equation}
\Omega=(\kappa\mu c R/2\pi\hbar)^3
\end{equation} where $\kappa < 1$ is some constant.   It follows
that
\begin{equation} {S\over  N\mu c R/\hbar}<{S\over 
N\Omega^{1/3}}={1\over
 N^{1/3}}{\ln(1+\bar n)+\bar n\ln(1+1/\bar n)\over \bar
n^{2/3}}
\end{equation} The function of $\bar n\equiv N/\Omega$ here
has a single maximum at
$\bar n=0.191$ with value $1.581$.  It follows that
\begin{equation} S<1.581 N \mu c R/N^{1/3}\hbar
\end{equation} Since $N>1$ it is obvious that the system satisfies
bound (\ref{bound1}) as required of a nonrelativistic assembly of
bosons.

It is plain that all we have shown must be true also for fermions:
the fermion phase space is more restricted than the boson one
(Pauli principle), so other things being equal, fermion entropy is
lower than boson entropy.   We have thus established the
correctness of the universal entropy bound for a collection of
nonrelativistic particles.   There is not much more to investigate
regarding the universal entropy bound, unless we turn to collections
of massless particles for which the more specific bound
(\ref{bound1}) is not relevant.   This we do now.

\section{The low temperature quandary}
\label{sec:temperature}

Deutsch~\cite{deutsch} originated the claim, which is occasionally
reinvented~\cite{FMW,wald,page1}, that a  system in a
thermal state violates the entropy bound~(\ref{bound}) if its
temperature $T=1/\beta$ is sufficiently low.  This is an
instructive issue.  It forces one to replace the definition of entropy
in Sec.~\ref{sec:mass} by that of entropy calculated according to
the canonical ensemble. 

Consider a system described by some massless quantum fields,
free or interacting, confined to a cavity of radius $R$.  I assume
there is a a unique  ground state of energy $\epsilon_0$.  For free
fields this assumption is trivial: the zero particles state is the
ground state.  For interacting fields it is a restrictive assumption
which I shall loosen up in Sec.~\ref{sec:field}.  I also assume there
is a
$g$-fold degenerate excited state at energy
$\epsilon_1=\epsilon_0+\Delta$, and higher energy states.  For
sufficiently large $\beta$ one may neglect the higher energy states
in the partition function
$Z=\sum_i\exp(-\beta\epsilon_i)$, and so approximate it by
$\ln Z\approx -\beta\epsilon_0 + \ln(1+g e^{-\beta\Delta})$. 
The {\it mean\/}  energy is
\begin{equation} E=-{\partial\ln Z\over
\partial\beta}=\epsilon_0+{g\Delta\over e^{\beta\Delta}+g},
\end{equation} while the entropy takes the form 
\begin{equation} S=\beta E+\ln Z={g\beta\Delta\over
e^{\beta\Delta}+g}+\ln(1+g e^{-\beta\Delta}) .\end{equation}
The typical claim is~\cite{deutsch}: ``measure energies from the
ground state so that $\epsilon_0=0$; then for the low
temperatures $\beta>2\pi R/\hbar c$ one gets $S>2\pi
RE/\hbar c$ and so bound~(\ref{bound}) is violated''. 

Early realistic numerical calculations of thermal quantum fields in
boxes~\cite{bek83} did reveal that,  were the ground state energy
to be ignored, bound~(\ref{bound}) would be violated at very low
temperatures, typically when 
$E<10^{-9}\hbar c/R$ ($R$ enters through the ``energy gap''
$\Delta$).  It was also clear early~\cite{bek81,bek83} that taking
any reasonable positive ground state energy into account
precludes the violation.  As the temperature rises, more and more
pure states are excited, and eventually $S/E$ peaks and begins to
decrease.  In this latter regime the entropy bound is always
obeyed regardless of whether or not one includes
$\epsilon_0$ in the total energy~\cite{bek84}.   I shall now
describe a proof of the result that the universal bound is
obeyed at arbitrarily low temperatures if the energy of all
components of the system is taken into account.

Taking the zero of energy of a system at its ground state is not
automatically justified because it may mean that $E$ in the
formulae is distinct from the gravitating energy.   But if the
particles involved are massless, what is the source of nonzero
$\epsilon_0$ ?   First of all some sort of boundaries must confine
the particles in the cavity. These should have some mass since they
must resist pressure of the particles.  In fact, their mass must be
positive on grounds of causality~\cite{bek82}.  In addition, those
boundaries will be responsible for a Casimir energy connected
with the particle species in question.   Although Casimir energies
can occasionally be negative~\cite{unwin}, the sum of boundary
and Casimir energies will be positive~\cite{bek94}.  But without 
going into details, we can state that $\epsilon_0$ must be larger
than $\hbar c/R$ because the system's Compton length must be
smaller than its size $R$ in order that the very notion of size  be
well defined.  For illustrative purposes I take $R\epsilon_0
/\hbar c > 2$.

The interesting quantity now is 
\begin{equation} S-2\pi RE/\hbar
c=\Xi(\beta\Delta)\equiv{(\beta\Delta-2\pi R\Delta)g\over
e^{\beta\Delta}+g}+\ln(1+g e^{-\beta\Delta})-2\pi
R\epsilon_0/\hbar c.\end{equation} The function $\Xi(y)$ is negative for
$y=0$ and $y\rightarrow\infty$, and has a single {\it maximum\/} at $y=2\pi
R\Delta$ where
$\Xi=\ln(1+g e^{-2\pi R\Delta/\hbar c})-2\pi
R\epsilon_0/\hbar c$. I thus conclude that
\begin{equation} S<2\pi RE/\hbar c+[\ln(1+g e^{-2\pi
R\Delta/\hbar c})-2\pi R\epsilon_0/\hbar c].
\label{last}
\end{equation} For the quantity in square brackets to be
nonnegative it would be necessary for $g\geq e^{2\pi
R\Delta/\hbar c}[e^{2\pi R\epsilon_0/\hbar c}-1]$, i.e.,
$g>2.87\times 10^5$. However, confined quantum field systems
do not exhibit such large degeneracy.  For example, for a free
scalar or electromagnetic field in a cubic cavity (which by virtue of
high symmetry should exhibit much degeneracy), $g$ is just a few
(there are a few lowest lying degenerate modes, and the lowest
excitation has one quantum in one of these modes)~\cite{bek84}.  
And a scalar field with a quartic self potential also exhibits little
degeneracy in its first excited levels~\cite{bek_guen}.  We thus see
why the square brackets in Eq.~(\ref{last})   are negative, so that
bound (\ref{bound}) is obeyed for our low temperature system of
massless quanta.  

\section{The high degeneracy quandary}
\label{sec:high}

The  argument in Sec.~\ref{sec:temperature} depends on the
supposition  that the degeneracy factor of  the lowest lying
excited states, $g$, cannot be large.   Although this is true in
many situations, it is not a law of nature.  It is possible to contrive
systems with large $g$.   Further elucidation of this issue is
possible by considering in some depth a purported
counterexample to the entropy bound by Page~\cite{page1}.

Page considers a sphere of radius $R$ partitioned into $n$
concentric shells; the partitions and the inner and outer
boundaries are regarded as infinitely conducting.  He points out
that the lowest ($\ell=1$) three magnetic-type electromagnetic
modes in the shell of median radius
$r$ have frequency $\omega\approx 1/r$.  Since there are $3n$
such modes (three for each shell), Page imagines populating now
one, then another and so on with a single photon of energy
$\sim \hbar c/r$ for the appropriate $r$.  These one-photon states
allow him to form a density matrix which, for equally weighted
states, gives entropy $\ln (3n)$ and mean energy $\sim 2\hbar
c/R$ (since $R/2$ is the median radius of the shells if they are
uniformly thick).  Page concludes that bound (1) is violated
because the entropy grows with $n$ while the mean energy does
not. 

Such an argument is wrong because it misses out part of the
energy.  The
$3n$ modes owe their existence to the infinitely conducting
partitions that confine them, each to its own shell.   To be highly
conducting, the envisaged partitions must contain a certain
number of charge carriers.  As we shall see, regardless of the
carriers' nature, their aggregated masses turn out to contribute enough
to the  system's total energy
$E$ to make it as large as required by the entropy bound
(\ref{bound}).  Ignoring the masses of the charge carriers goes
against the condition that the bound applies to a complete system:
the carriers are an essential component, so their gravitating energy
has to included in $E$.  The situation must be contrasted with that
in which the electromagnetic field is  confined to an
\emph{empty} sphere (or for that matter any empty
parallelepiped).   Detailed state counting~\cite{bek84} has shown
that the entropy bound is satisfied even if one leaves out the
energy contribution from outside the photon system (walls of
confining cavity).  In the case of Page's rather contrived onion-like
system, the energy of the supporting structures cannot be ignored; its
inclusion in the bound is justified by the above remarks.

I assume all partitions to have equal thickness $d$.  One
mechanism that can block the waves from crossing a partition is a
high plasma frequency
$\omega_p$ of the charge carriers in the partitions (which I do
\emph{not} assume to be electrons necessarily).  We
know~\cite{jackson} that in a plasma model of a conductor with
collisionless charge carriers, the electromagnetic wave vector for
frequency
$\omega$ is
$k=\omega c^{-1}(1-\omega_p^2/\omega^2)^{1/2}$, so that if
$\omega<\omega_p$, the fields do not propagate. 
Nevertheless they do penetrate a distance
$\delta=c\omega^{-1}(\omega_p^2/\omega^2-1)^{-1/2}>c\omega_p{}^{-1}$
into the plasma before their amplitudes become insignificant.  In
order to prevent these evanescent waves from bridging a
partition, one must thus require
$\delta<d$, i.e., $\omega_p d>c$.  But 
\begin{equation}
\omega_p{}^2=4\pi{\cal N}e^2/m,
\label{plasma}
\end{equation} where ${\cal N}$ is the density of charge carriers
of charge
$e$ and mass $m$.  Since $d<R/n$, all this gives us
$(4\pi R^2 d){\cal N}>m c^2 n^2 d/e^2$.  Now $4\pi R^2 d$ is
the volume of material in the outermost partition.  Properly
accounting for the variation of partition area with its order $i$ in
the sequence (we employ the sum
$\sum i^2$),  tells us that for $n\gg 1$ the total mass-energy in
charge carriers in all the partitions is $E\approx nmc^2(4\pi R^2
d/3){\cal N}$.  Substituting our previous bound on $(4\pi R^2
d){\cal N}$, I get
$E>{\scriptstyle 1\over
\scriptstyle 3}n^3 m^2 c^4 d/e^2$.  

Now as a matter of principle $e^2<\hbar c$ (recall
that in our world $e^2\approx \hbar c/137$), because more
strongly coupled electrodynamics would make structures, e.g.
atoms and partitions, which are all held together electrically,
unstable~\cite{rozental}.  We also evidently have $R> nd$.  Hence
$ER/\hbar c>{\scriptstyle 1\over
\scriptstyle 3}n^4(mcd/\hbar)^2$.   But a charge carrier's Compton length
has to be smaller than $d$, for otherwise the carriers would not be
confined to the partitions; thus
$m c d>\hbar $.  Hence $2\pi RE/\hbar c>2
n^4$ which is always  larger than the  entropy in photons $\ln
(3n)$. 

The only alternative mechanism for keeping electromagnetic
waves from penetrating into a conductor is the skin
effect~\cite{jackson}.  The skin depth for electromagnetic waves of
frequency $\omega$ is
$\delta_s\approx c(2\pi\omega\sigma)^{-1/2}$, where
$\sigma$ is the conductivity.  In the simple Drude
model~\cite{jackson},
$\sigma={\cal N}e^2(m/\tau-\imath m\omega)^{-1}$, where
$\tau$ is a charge carrier's slowing-down timescale  due to
collisions, and
$\imath=\sqrt{-1}$.  The $\sigma$ in the expression for
$\delta_s$   refers to an Ohmic (real) conductivity rather than to an
inductive (imaginary) one.  Thus one must demand that
$\omega\ll 1/\tau$.  This is no real restriction since one is
interested in photons with the lowest possible energy.   But then 
\begin{equation}
\delta_s\gg (2\pi{\cal N}e^2/mc^2)^{-1/2}.
\label{skin}
\end{equation} As before one must require $\delta_s < d<R/n$. 
This gives 
$(4\pi R^2 d){\cal N}\gg 2mc^2 n^2 d/e^2$ which is just a
stronger version of the lower bound on ${\cal N}$ we got before.  
Repeating the previous discussion {\it verbatim\/} shows that
$2\pi RE\gg 4n^4$, which bounds the photon's entropy $\ln
(3n)$ confortably.

Of course, the charge carriers (and the lattice through which they
move) also contribute to the entropy.   However, they constitute a
nonrelativistic system, and for such the results of
Sec.~\ref{sec:mass} assure us that the universal entropy bound is
obeyed, with much room to spare.   What we have thus just shown is that
adding the entropy of the photons will not change the situation.  
Although in Page's onion structure the
photons by themselves may violate bound (\ref{bound}), this
bound is satisfied by the complete system, photons $+$ charge
carriers.

\section{The low excitations quandary}
\label{sec:excitation}

In the framework of the microcanonical ensemble there is a potential
challenge to bound (\ref{bound}) if the energy gap between the ground state
and the first excitations is very small.   We illustrate this with Page's'
example~\cite{page2}  of  the electromagnetic field confined to a
coaxial cable of length $L$ which is coiled up so as to fit within a
sphere of radius
$R$, with $R\ll L$,  before being connected end to end to form a
loop.   

Page's entirely qualitative reasoning proceeds by analogy with a
rectilinear coaxial cable with periodic boundary conditions. A
rectilinear infinitely long coaxial cable has some electromagnetic
modes which propagate along its axis with arbitrarily low
frequency.  Page notes that for the coiled-up cable, each right
moving mode is accompanied by a degenerate (in frequency) left
moving mode (basically this follows from time reversal invariance
of Maxwell's equations).  He then argues that if the cable's {\it
outer\/} radius $\rho_2$ is small on scale $R$, the structure of the
electromagnetic modes is little affected by the cable's curvature.
This leads him to estimate the lowest frequency $\omega_1$ as
similar to that of the rectilinear coaxial cable with periodic
boundary conditions with period
$L$: $\omega_1\approx 2\pi c L^{-1}$.  Page notes that there are
three electromagnetic states with energies $\Delta\equiv
E-E_{\rm vac}\leq 2\pi \hbar c L^{-1}$: the vacuum, and a single
photon in the right- or in the left-moving mode of frequency
$\omega_1$.  Therefore, up to energy
$\Delta$ above the vacuum, there is entropy $S=\ln 3\approx 1$
.  Since
$2\pi R\Delta /\hbar c\approx 4\pi^2 (R/L)$, which could be
very small compared to unity,  Page points to this example as a
violation of the entropy bound.

By now we are experienced enough to see where the error lies.  The
interesting question is rather whether the  electromagnetic field
{\it plus\/} coaxial cable (complete system) complies with 
bound~(\ref{bound}). Now the inner conductor of the cable---let
its radius be
$\rho_1$---is an essential part of the system, for without it the
lowest propagating frequency would be
$\omega_1\sim  c\rho_2{}^{-1}$, where $\rho_2$ is the outer
radius of the cable, and thus very large on scale
$c L^{-1}$.  To play its role, the inner conductor must keep the
electromagnetic fields out of it against the two mechanisms which
permit such fields to penetrate into a conductor.  I now elaborate
on this.  

As mentioned in Sec.~\ref{sec:high}, a low frequency electromagnetic wave 
penetrates into a conductor a distance
$\delta> c\omega_p{}^{-1}$ with the plasma frequency $\omega_p$ given
by Eq.~(\ref{plasma})  (again ${\cal N}$ here is the density of charge carriers
of charge $e$ and mass $m$).  In order to prevent these evanescent waves
from bridging the inner conductor, one must require
$\rho_1>\delta$  so $\rho_1>c\omega_p$
(I assume a solid inner
conductor).  Then by Eq.~(\ref{plasma})
${\cal N}\rho_1{}^2>mc^2(4\pi e^2)^{-1}$.  Now the volume of
material in the inner conductor is $\pi\rho_1{}^2 L$, so that its
mass-energy is at least 
$\pi\rho_1{}^2 L{\cal N}mc^2$; thus the total mass-energy
$E$ of cable plus field is  constrained by $E>m^2c^4 L(4
e^2)^{-1}$.  However, for the charge carriers to be localized within
the conductor, their Compton lengths must be smaller than
$\rho_1$, so that $mc\rho_1\gg \hbar$.  Hence
$E\gg \hbar^2 c^2L(4 e^2 \rho_1{}^2)^{-1}$, or
\begin{equation}  2\pi ER/\hbar c\gg(\pi\hbar
c/2e^2)(L/\rho_1)(R/\rho_1).
\label{ratio}
\end{equation}
 Since $e^2<\hbar c$  (see Sec.~\ref{sec:high}), and since $L\gg
R>\rho_1$ by the conditions of the problem,
$2\pi ER/\hbar c\gg 1$, and thus bound~(\ref{bound})
confortably bounds the photons' entropy
$S=\ln 3$.  

However, even when $\rho_1>\delta$ is satisfied, the
waves may bridge the inner conductor if its skin depth
[Eq.~(\ref{skin})] approaches $\rho_1$.  
${\cal N}\rho_1{}^2\gg mc^2(2\pi e^2)^{-1}$.  But this is just a
stronger version of our earlier lower bound on ${\cal N}$.  
Repeating the previous discussion shows again that
$2\pi ER/\hbar c$, with $E$ the total energy of the system, again
bounds the photons' entropy $S=\ln 3$ with plenty of room to spare.

In putting a lower bound on $E$, I ignored the cable's outer
conductor (positive energy) and the Casimir energy.  If positive
this last only makes the case for the bound stronger.  What if it is
negative ? As a rule~\cite{unwin} the magnitude of the
electromagnetic Casimir energy of a cavity is a small fraction
($10^{-3}$ to $10^{-2}$) of  $\hbar$ times the lowest
eigenfrequency, here $\omega_1\approx 2\pi c L^{-1}$.  Because
$e^2<\hbar c$ and $\rho_1\ll L$, the lower bound on $E$
recorded just prior to Eq.~(\ref{ratio}) is vastly greater than the Casimir
energy, which may thus be neglected.  

One might think that this success of the entropy bound hinges on
the rather low entropy Page associated with the system.  So
suppose one enlists all low-lying  electromagnetic modes devoid
of transversal nodes and having wavelengths along the cable's axis
of the form $L/k$ with
$k=1,2,3,\,\cdots$.  There are $\tilde N=\mathcal{O}(2L/\rho_2)$ such
doubly degenerate modes with frequency below $c/\rho_2$,  the
frequency of the lowest-lying transversally excited mode which
itself serves as base for a separate, second series of modes. How
many states can one build from these
$\tilde N$ modes, states whose energies lie below $\hbar
c/\rho_2$, the energy of the lowest state arising from the second
series of modes ?  There are obviously $\tilde N$ one-photon
states, fewer than
$\tilde N^2/2!$ two-photon states, fewer than  $\tilde N^3/3!$
three-photon states, etc. (recall: photons are indistinguishable). 
Together with the vacuum's contribution of unity,  the series of
upper bounds sums to $e^{\tilde N}$.  Taking the logarithm we
have for the entropy
$S<\mathcal{O}(2L/\rho_2)<(L/\rho_2)(R/\rho_2)$.  But  by
inequality~(\ref{ratio}) together with the conditions $e^2<\hbar
c$ and
$\rho_1<\rho_2$, this last factor is bounded from above by $2\pi
ER/\hbar c$.  The futility of trying to violate
bound~(\ref{bound}) if $E$ includes the cable's mass-energy is
thus clear.   Of course, I have ignored the entropy contributed by
the charge carriers.   As argued already in Sec.~\ref{sec:high}, any
such entropy is bounded,  with room to spare,  by inequality
(\ref{bound1}) just on the basis of the carriers' rest masses. 
Therefore, we can afford to focus on the photon entropy alone.  To
sum up, {\it whatever\/} the construction of the coaxial cable, the
whole system complies with bound~(\ref{bound}) as long as $E$
includes the cable's mass energy. 

Page felt justified in ignoring the energy associated with the cable 
in view of a theorem by Schiffer and me~\cite{schiffer,review} to
the effect that a massless free scalar field confined to a cavity of
whatever topology by Dirichlet boundary conditions obeys bound
(\ref{bound}), even if
$E$ is taken as the energy above the vacuum state.  This would
seem to condone the interpretation of $E$ in bound (\ref{bound})
as energy above the ground state.  The electromagnetic field is
another story.  In Ref.~\cite{schiffer} we sketched how the
theorem could be extended to the electromagnetic and other
noninteracting fields, but this generalization was never formally
proved.  In our review, Ref.~\cite{review}, we work only with the
scalar case.  \emph{If} Page's estimate of
$\omega_1$ is correct,  such a theorem cannot apply to the
electromagnetic field in a cavity with not simply connected
crossection, like the coaxial cable.

\section{The degenerate ground state quandary}
\label{sec:degenerate}

Confined free fields have a unique ground state, the vacuum.  But
nonlinear fields can have multiple degenerate vacua.  For
instance, the scalar field with a double well self-potential has two
classically degenerate ground states, each with the field locked
everywhere at the minimum of one of the wells.  If the potential
minima are zero, there are two states at zero energy.    Although
then $S=\log 2>0$, this is not a counterexample to bound
(\ref{bound}) since $2\pi RE/\hbar c$ is indefinite because
$R$ is formally infinite (the field has a constant nonzero value
everywhere).

\subsection{Nonlinear scalar field}
\label{sec:field}

Page~\cite{page2} purports to construct a real counterexample
by considering configurations in which the said scalar field
vanishes at a certain boundary of radius $R$.  He correctly
points out that the aforesaid classical degenerate ground states
engender, by quantum tunnelling between the wells, a new
ground state $\psi_0$ (energy $\epsilon_0$) with equal
amplitude at each well and a first excited state $\psi_1$ a very
small energy $\epsilon_1-\epsilon_0$ above the ground state.
Since the entropy of a mixed state containing $\psi_0$ and
$\psi_1$ can reach $\ln 2$ (ground and excited states equally
probable), while
$(\epsilon_1-\epsilon_0)R/\hbar c$ can be exponentially small 
for deep wells, Page was convinced that the described state
violates bound (\ref{bound}). 

Page identifies the energy $E$ of bound~(\ref{bound}) with 
$\epsilon_1-\epsilon_0$, the energy measured {\it above\/} the
ground state.  This would be correct if the ground state referred to
a spatially unrestricted configuration, because then the bottom of
a potential well would be the correct zero of energy (neglecting
zero point fluctuations).   But since the field is required to vanish
at radius $R$, the energy
$\epsilon_0$ of the described ground state is a function of
$R$, and it makes little sense to take it as the zero of energy.  For
example, by expanding the system can do work ($-\partial
E/\partial R\neq 0$), so that its gravitating energy changes, and
cannot be taken as zero for all $R$.  The gravitating energy for the
equally likely mixture of ground and excited states should be
identified with
${\scriptstyle 1\over \scriptstyle
2}(\epsilon_0(R)+\epsilon_1(R))\approx
\epsilon_0(R)$. The exponential smallness of
$\epsilon_1-\epsilon_0$ is not
very relevant for the issue of the bound's validity as I will show.

Since there are no solitons in $D=3+1$ spacetime~\cite{derrick}, a
finite sized field configuration [the only interesting
case---see~(\ref{bound})] must be confined by a ``wall'' which
cannot be ignored, as Page does, if we stick to the original form of
the entropy bound.  There are three parts to the energy $E$ of the
complete system (before tunnelling is taken into account): the
classical energy $\epsilon_c$ of the field configuration
concentrated around one well but vanishing at radial coordinate
$r=R$, the quantum correction
$\epsilon_v$ due to the zero point fluctuations about the classical
configuration, and $\epsilon_w$, the energy of the ``wall'' at
$r=R$.  As I show below, $\epsilon_w$ is at least of the same
order as $\epsilon_c$, and both strongly dominate the energy
$\epsilon_1-\epsilon_0$.

\subsection{Classical two-well configurations}
\label{two}

The double well potential field theory comes from the lagrangian
density
\begin{equation}    {\cal L}=-\hbar c\Big[{1\over
2}\partial_\mu\phi\,\partial^\mu\phi+{1\over
4}\lambda(\phi^2-\phi_m^2)^2\Big]
\label{theory} .
\end{equation} This gives  the field equation
\begin{equation}  
\partial_\mu\,\partial^\mu\phi-\lambda\phi(\phi^2-\phi_m^2)=0
\label{TD}.
\end{equation} Every spherically symmetric configuration inside a
spherical box of radius $R$ will thus satisfy (I use standard
spherical coordinates;
$'$ denotes derivative w.r.t. to $r$)
\begin{equation}  
r^{-2}(r^2\phi')'-\lambda\phi(\phi^2-\phi_m^2)=0
\label{TI} .\end{equation} Regularity requires that $\phi'=0$ at
$r=0$.  Page chooses $\phi=0$ at
$r=R$.  The classical energy of such a configuration will be
\begin{equation}   \epsilon_c={\hbar c\over
 2}\int_0^R\left[\phi'^2+{\scriptstyle 1\over \scriptstyle
2}\lambda(\phi^2-\phi_m^2)^2\right] r^2\,dr
\label{Ec} .\end{equation}

Since one is interested in the ground state, I require that
$\phi$ have its first zero at $r=R$.  Multiplying Eq.~(\ref{TI}) by
$r^2\phi$ and integrating over the box allows, after integration
by parts and use of the boundary conditions, to show that
\begin{equation}  
\int_0^R\phi'^2r^2\,dr=\lambda
\int_0^R(\phi_m^2-\phi^2)\phi^2 r^2\,dr
\end{equation} whereby
\begin{equation}   \epsilon_c={ \lambda\hbar c\over 
4}\int_0^R(\phi_m^4-\phi^4)r^2\,dr
\label{energy} .\end{equation}

It proves convenient to adopt a new, dimensionless, coordinate
$x \equiv \surd\lambda\ \phi_m\,r$ and a dimensionless scalar
$\Phi\equiv
\phi/\phi_m$.  Then Eq.~(\ref{TI}) turns into a parameter-less
equation:
\begin{equation}   {1\over x^2}{d\over dx}\left(x^2{d\Phi\over
dx}\right)+\Phi(1-\Phi^2)=0
\label{TI2} .\end{equation} Using $d\Phi/dx=0$ at $x=0$ one
may integrate the equation to get
\begin{equation}   {d\Phi\over dx}=-{1\over
x^2}\int_0^x\Phi(1-\Phi^2)\,x^2\,dx.
\label{helpful}
\end{equation} If the integration starts with $\Phi(0)>1$, then by
continuity the r.h.s. of Eq.~(\ref{helpful}) is positive for small $x$,
so that
$\Phi$ grows.  There is thus no way for the r.h.s. to switch sign, so
$\Phi(x)$ is monotonically increasing and can never have a zero. 
If
$\Phi(0)=1$, it is obvious that the solution of Eq.~(\ref{helpful}) is
$\Phi(x)\equiv 1$ which cannot satisfy the boundary condition at
$r=R$.   Thus the classical ground state configuration we are after
requires
$\Phi(0)<1$.

When $\Phi(0)<1$ it can also be seen from Eq.~(\ref{helpful})
that $\Phi$ is monotonically decreasing with $x$.  For a particular
$\Phi(0)$, $\Phi(x)$ will reach its first zero at a particular $x$
which I refer to as $x_0$.  This can serve as the parameter singling
out the solution in lieu of
$\Phi(0)$.  One thus has a family of ground state configurations
$\Phi(x,x_0)$.  Each such configuration corresponds to a box of
radius
$R=x_0(\surd\lambda\ \phi_m)^{-1}$.  In terms of the new
variables one can write Eq.~(\ref{energy}) as
\begin{equation} \epsilon_c ={\hbar cx_0\over 4\lambda
R}\int_0^{x_0}(1-\Phi^4)x^2\, dx.
\label{Ec2}
\end{equation} The dependence  $\epsilon_c\propto
\lambda^{-1}$ is well known from kink solutions of
theory~(\ref{theory}) in
$D=1+1$~\cite{rajaraman}, where the role of $\hbar /c R$ is
played by the effective mass of the field.   Numerical integration of
Eq.~(\ref{TI2}) shows that the factor
$x_0\int_0^{x_0}(1-\Phi^4)x^2\, dx$ grows monotonically from
32.47 for
$\Phi(0)= 0\ (x_0=3.1416)$ to 232.23 for
$\Phi(0)=0.98\ (x_0=5.45)$ to infinity as $\Phi(0)\rightarrow 1\
(x_0\rightarrow\infty)$. Since $\epsilon_c$ is not exponentially
small, the quantum tunnelling corrections that Page discussed are
negligible, so one need only  add to $\epsilon_c$ the zero point
fluctuations energy
$\epsilon_v$ plus the wall energy
$\epsilon_w$ to get the full energy associated with the ground
state.  This sum plays the role of $E$ in the bound (\ref{bound}). 

I shall not bother to calculate $\epsilon_v$ (which should include
the Casimir energy).  This can be done by present techniques only
for the weak coupling case $\lambda<1$~\cite{rajaraman}.  It is
then found in other circumstances, e.g. the $D=1+1$ kink, that
$\epsilon_v$ is small compared to $\epsilon_c$.  The situation for
large
$\lambda$ (the strong coupling regime) is unclear.  However, it is
appropriate to recall here that the theory (\ref{theory}) is trivial in
that it makes true mathematical sense only in the case
$\lambda=0$~\cite{callaway}.  Theorists use it for
$\lambda\neq 0$ to obtain insights which are probably
trustworthy in the small $\lambda$ regime, but probably not for
large $\lambda$.

I now set a lower bound on $\epsilon_w$.  A look at
Eq.~(\ref{Ec2}) shows that for
$\Phi(0)\ll 1$ and so $\Phi(x)\ll 1 $), $\epsilon_c$ scales as
$x_0^4/R\propto R^3$. Numerically the exponent of $R$ here
only drops a little as $\Phi(0)$ increases; for example, it is 2.86 for
$\Phi(0)=0.98$.  So I take it as 3 for now.  On virtual work grounds
(consider expanding $R$ slightly), the $R^3$ dependence means
the
$\phi$ field exerts a suction (negative pressure) of dimension
$\approx (3\epsilon_c/4\pi R^3)$ on the inner side of the wall. 
By examining the force balance on a small cap of the wall, one sees
that in order for the wall to withstand the negative pressure, it
must support an internal compression (force per unit length)
$\tau\approx (3\epsilon_c/8\pi R^2)$~\cite{bek82}.  Under this
compression vibrations on the wall will propagate superluminally
unless the surface energy density is at least as big as $\tau$
(dominant energy condition).  Thus one may conclude that the
wall (area
$4\pi R^2$) must have (positive) energy $\epsilon_w>
3\epsilon_c/2$ which {\it adds\/} to $\epsilon_c$ to give
$E>5\epsilon_c/2$.   

As mentioned, for $\Phi(0)$ very close to unity the exponent $n$
in
$\epsilon_c\propto R^n$ falls below 3; as a consequence the
coefficient in the previous inequality is somewhat lower than
$5/2$.  However, by then
$\epsilon_c R$ is already much larger than the corresponding
quantity for
$\Phi(0)\ll 1$  (six times larger for $\Phi(0)=0.98$). Using the
value of
$\epsilon_c$ for
$\Phi(0)\ll 1$ from the preceding argument, I thus conclude that
for all physically relevant  $\Phi(0)$, $2\pi ER/\hbar c >
127.5\lambda^{-1}$.  This is certainly not exponentially small as
Page originally claimed !  

True, formally  it seems possible to have a violation of the bound 
for the 50\% mixture of ground and excited states ($S=\ln 2$)
whenever
$\lambda> 127.5/\ln 2 =183.95$.  However, this is the strong
coupling regime.  For all one knows the zero point energy
$\epsilon_v$ may then become important and tip the scales in
favor of the entropy bound.  At any rate, because the theory
(\ref{theory}) is trivial, one is more likely to be overstepping here
the bounds of its applicability than to be witnessing a violation of
the entropy bound at large $\lambda$.   Indeed, in $D=1+1$
spacetime Guendelman and I~\cite{bek_guen} found analytically
all static classical configurations for the interacting theory
(\ref{theory}) in a box, and their energies {\it sans\/} the box's. 
The distribution of energy levels turns out to be such that the
entropy bound~(\ref{bound}) is sustained.  But I know of no
analogous result in
$3+1$ dimensions

In summary, I have shown that whenever the calculation is
meaningful ($\lambda$ not large), the entropy bound
(\ref{bound}) is satisfied in Page's example provided $E$ includes
all contributions to the energy.  Page~\cite{page2} does not
disagree with this finding, but he cites my paper with
Schiffer~\cite{schiffer} as an excuse for including in
$E$ just the excitation energy above the classical ground state. 
However, we ourselves restricted use of this approach to an
assembly of quanta of a massless {\it noninteracting\/} field;  
the  theorem for scalar fields I mentioned in
Sec.~\ref{sec:excitation} is of no help here because it holds for
noninteracting fields only.     Although testing the bound by
ignoring the ground state energy~\cite{bek84,review} is rather
straightforward, it should not make us forget that in the
universal bound, $E$ must include the ground state energy.  
  
\subsection{Multiwell potential}
\label{multi}

Page~\cite{page1} also confronts bound (\ref{bound}) with a
theory like  (\ref{theory}) but with a potential having three
equivalent wells. Pressumably one would like one of these
centered at $\phi=0$, with the other two flanking it
symmetrically.  Then Page's conclusion that there are three
exponentially close states (in energy) is untenable.  This would
require three classically degenerate configurations, which
certainly exist in open space (field $\phi$ fixed at one of three well
bottoms).  However, one is here considering a finite region with
$\phi=0$ on the boundary  $r=R$.  One exact solution is indeed
$\phi\equiv 0$, and it has zero energy (the zero point fluctuation
energy correction will, however, depend on
$R$).  Then there are two degenerate solutions in which the field
starts at
$r=0$ in one side well and then moves to the central one with
$\phi\rightarrow 0$ as
$r\rightarrow R$.  By analogy with our earlier calculations, the
common classical energy of these two configurations will be of 
$\mathcal{O}(8\hbar c/\lambda R)$.  It cannot thus be regarded as the zero
of energy; this role falls to the energy of the $\phi=0$
configuration.  When tunnelling between wells is taken into
account, one has a truly unique ground state and two excited
states of classical origin split slightly in energy (plus the usual
gamut of quantum excitations).  The entropy of an equally
weighted mixture of these states is $\ln 3$.  Ignoring Casimir
energy, its  mean energy
$E$ is ${\scriptstyle 2\over
\scriptstyle 3}(\epsilon_c+\epsilon_ w)$ of an excited state, that is
$\mathcal{O}(13\hbar c/\lambda R)$, so the entropy bound is easily
satisfied, at least in the weak coupling regime where the theory
makes sense.

When the potential has $n=5,7,9,\cdots\ $ equivalent wells with
one centered at $\phi=0$ and the rest disposed symmetrically
about it, there will be a single zero-energy configuration
($\phi\equiv 0$), and
${\scriptstyle 1\over \scriptstyle 2}(n-1)$ pairs of degenerate
configurations with succesively ascending $R$-dependent
energies.  For
$n=4,6,8,\cdots\ $ wells there is no zero-energy configuration, but
there are
${\scriptstyle 1\over \scriptstyle 2}n$ pairs of degenerate
configurations with $R$ dependent energies.  Because of the extra
energy splitting appearing here already classically, I expect by
analogy with the previous results,  that the appropriate mean
configuration energy (perhaps supplemented by wall energy),
when multiplied by $2\pi R/\hbar c$, will bound the maximum
entropy, $\ln n$, from above.

\subsection{Zero mode systems}
\label{zero}

Years ago Unruh~\cite{unruh} proposed a counterexample to the
universal entropy bound which has some resemblance to those based
on degenerate ground states.  He focused on a system with a zero
(frequency) mode.  For example, a real massless scalar field confined
to a box by Neumann boundary conditions has a zero frequency mode:
$\phi=q+pt$ with $q$ and $p$ real constants.  Unruh
noted that a vacuum state of such a scalar field is
exclusively characterized by its
$q$ and $p$.  Noting further that one can add an arbitrary constant to
$q$, he asserted that the system has an infinite number of
degenerate vacua, with a common energy determined by $p^2$. 
This would evidently violate bound (\ref{bound}) by an infinite
factor.

The problem with this proposal~\cite{BS} is the identification of 
zero modes with distinct $q$ with one and the same system.  This
might make sense if the zero mode could be populated with quanta
just as a
$\omega\neq 0$ mode can.  But in fact, a zero mode represents the
classical part of $\phi$.  It does not represent quanta: different
$q$ do not stand for different ``occupation numbers'' in the ground
state, but for different systems.  The $q$ is very much like the order
parameter in a superfluid which tells us the density
of particles in the superfluid ground state, which in turn serves as base
for all other, excited, states.    Changing the order parameter
defines a different superfluid (different particle density). Nobody
would think of counting all superfluid systems to compute an entropy
(which would of necessity be infinite).  Likewise, it makes no sense to
ascribe entropy to the multiplicity of $q$ values.

\section{The proliferation of species quandary}
\label{proliferation}

A popular challenge to the entropy
bound~\cite{UW,FMW,page1} imagines an hypothetical
proliferation of particle species.  Suppose there were to exist as many
copies $\tilde N$ of a field e.g. the electromagnetic one, as one
ordered.  The entropy in a box containing a fixed energy allocated to
the said fields should grow with $\tilde N$ because the bigger $\tilde
N$, the more ways there are to split up the energy.  Thus eventually
the entropy should surpass the entropy bound.  Numerical estimates
show that it would take
$\tilde N\sim 10^9$ to do the trick~\cite{bek83}. A similar picture
seems to come from Eq.~(\ref{last}); the degeneracy factor $g$
should grow proportionally to $\tilde N$ making the factor in square
brackets large so that, it would seem, one could not use the
argument based on (\ref{last}) to establish that $S<2\pi RE/\hbar
c$.    These observations constitute the quandary of the
proliferation of species.

There are several approaches to its resolution.  As already remarked in
Ref.~\cite{bek82} and reiterated later~\cite{bek94,bek99}, the above
reasoning fails to take into account that each field species contributes
to the Casimir energy which gets lumped in $\epsilon_0$.  If these
contributions are positive, then the negative term in the square bracket
in Eq.~(\ref{last}) eventually dominates the logarithm as $\tilde N$
grows, and for large$\tilde N$ one again recovers the entropy
bound~(\ref{bound}).  If they are negative (which implies a Casimir
suction proportional to $\tilde N$ on the walls which delineate the
system), then the scalar field example suggests that the wall energy,
which must properly be included in $\epsilon_0$, should suffice to
make the overall $\epsilon_0$ positive~\cite{bek94}.  This would go
a long way towards making the entropy bound safe.  

Bousso~\cite{bousso04} has pointed out that when some species 
contribute positive Casimir energies and some contribute negatively, a
near cancellation of the Casimir energy could take place making the
above saving strategy irrelevant.  He suggests that radiative
corrections to the interactions responsible for confining the particles
of interest in the cavity make a contribution to $\epsilon_0$
proportional to $\tilde N$ which suffices to make the entropy bound
work.

There is an alternative view~\cite{bek82-2,bek99}: the seeming
clash between entropy bound and an exceedingly large number of species
merely tells us that physics is consistent only in a world with a
limited number of particle species, such as ours. 
Indeed, as Brustein, Eichler, Foffa and Oaknin have
argued~\cite{BEF}, a very large number of species will make
the vacuum of quantum field theory unstable against collapse into a
``black hole slush'', unless we are willing to accept a rather modest
ultraviolet cutoff for the theory.  The unlimited proliferation of species 
may not even be physically consistent, and cannot thus constitute
an argument against the entropy bound.
 
\acknowledgments  This research was supported by the Israel 
Science Foundation under grant  129/00.

\end{document}